\documentclass[manuscript]{aastex}
\usepackage{rotating}

\begin{document}


\title{$HST$ Fine Guidance Sensor Parallaxes for Four Classical Novae$^{\rm 1,2}$}

\author{Thomas E. Harrison$^{\rm 3}$, Jillian Bornak}

\affil{Department of Astronomy, New Mexico State University, Box 30001, MSC 
4500, Las Cruces, NM 88003-8001}

\email{tharriso@nmsu.edu, jbornak@nmsu.edu}

\author{Barbara E. McArthur and G. Fritz Benedict}

\affil{McDonald Observatory, University of Texas, Austin, TX 78712}

\email{mca@barney.as.utexas.edu, fritz@astro.as.utexas.edu}

\begin{abstract}
We have used data obtained with the Fine Guidance Sensors on the {\it Hubble Space 
Telescope} to derive precise astrometric parallaxes for four classical novae: V603 
Aql, DQ Her, GK Per, and RR Pic. All four objects exceeded the Eddington limit at 
visual maximum.  Re-examination of the original light curve data for V603 Aql and
GK Per has led us to conclude that their visual maxima were slightly brighter than 
commonly assumed. With known distances, we examine the various 
maximum magnitude$-$rate of decline (MMRD) relationships that have been established 
for classical novae. We find that these four objects show a similar level of
scatter about these relationships as seen in larger samples of novae whose 
distances were determined using indirect techniques. We also examine the nebular
expansion parallax method, and find that it fails for three of the four objects.
In each case it was possible to find an explanation for the failure of that
technique to give precise distance estimates. DQ Her appears to suffer from
an anomalously high extinction when compared to field stars on its sight line.
We suggest that this is likely due to local material, which may also be the
source of the $IRAS$ detections of this object.
\end{abstract}

\noindent
{\it Key words:} parallaxes --- stars: novae, cataclysmic variables --- stars: 
individual (V603 Aquilae, DQ Herculis, GK Persei, RR Pictoris)

\begin{flushleft}
$^{\rm 1}$Based partially on observations made with the NASA/ESA Hubble Space 
Telescope, obtained from the Data Archive at the Space Telescope Science Institute, 
which is operated by the Association of Universities for Research in Astronomy, 
Inc., under NASA contract NAS 5-26555. These observations are associated with 
programs GO10912, GO11295, and GO11785.\\
$^{\rm 2}$Based partially on observations obtained with the Apache Point 
Observatory 3.5-meter telescope, which is owned and operated by the Astrophysical 
Research Consortium.\\
$^{\rm 3}$Visiting Astronomer, Cerro Tololo Inter-American Observatory, National
Optical Astronomy Observatory, which is operated by the Association of 
Universities for Research in Astronomy, Inc., under cooperative agreement with
the National Science Foundation.\\
\end{flushleft}
\clearpage

\section{Introduction}

Classical novae (CNe) are thermonuclear explosions on the surface of a white 
dwarf that has been accreting material for thousands of years from its low mass 
companion. Townsley \& Bildsten (2004) show that for cataclysmic variable
systems with mass accretion rates of $\dot{M}$ = 10$^{\rm -8}$ to 
10$^{\rm -10}$ M$_{\sun}$ yr$^{\rm -1}$, CNe ignition can occur once the accumulated 
envelope on the white dwarf reaches 10$^{\rm -4}$  to 10$^{\rm -5}$ M$_{\sun}$.
The resulting eruption can reach well 
beyond the Eddington limit, and eject $\approx$ 10$^{\rm -4}$ 
M$_{\sun}$ of enriched material at high velocity. With the most luminous
eruptions reaching to M$_{\rm V} \leq$ $-$9.0, and due to the fact that they 
occur in all types of galaxies, CNe have been proposed as useful extragalactic 
distance indicators (van den Bergh \& Pritchet 1986; Della Valle \& Livio 1995; 
Della Valle \& Gilmozzi 2002). In addition, however, the eruptions of CNe
provide critical tests of our understanding of thermonuclear runaways, 
the nucleosynthesis that occurs within the burning layers (c.f., Starrfield
et al. 2009), and the factors that drive and shape the shell ejection process.

To fully understand the outbursts of CNe, it is essential to have precise
distances. While a wide range of secondary distance estimation techniques have
been applied to CNe, none have had high precision parallaxes measured. The
most reliable indirect method for estimating CNe distances has come from ``nebular
expansion parallaxes''. This technique combines spectroscopically 
determined expansion velocities, and the observed nebular remnant shell size, to 
estimate the distance.
For the earliest attempts to employ this technique, the velocity of the 
``principal absorption component'' (see Payne-Gaposchkin 1957, or Warner 2008) was 
used to estimate
the expansion velocity of the bulk of the ejecta. Unfortunately, such spectra
are only seen near visual maximum, and thus rarely observed for most 
CNe. More recent efforts (see the review by O'Brien \& Bode 2008) employ
a kinematic model derived from spectroscopic observations of the resolved
shell. This regimen is much more robust, in that it allows compensation for
the tendency of CNe to have prolate, ellipsoidal remnants (Wade et al. 2000).

One of the long-standing correlations in the field of CNe, dating back to McLaughlin
(1945), is that the {\it speed} of the outburst is related to the peak 
luminosity of the eruption. McLaughlin used a variety of distance
estimation techniques to derive the absolute visual magnitudes at maximum,
and correlated this with the time it took for the CNe to dim by three magnitudes
from visual maximum (``$t_{\rm 3}$''). There have been a number of attempts to 
calibrate a maximum magnitude$-$rate of decline (MMRD) relationship for CNe.
Downes \& Duerbeck (2000) have produced the most recent updates
(though see Hachisu \& Kato 2010), including the two linear laws (involving 
$t_{\rm 2}$ and $t_{\rm 3}$), as well as the arctangent law 
(that uses $t_{\rm 2}$) first formulated by Della Valle \& Livio (1995).
The conclusion of Downes \& Duerbeck was that a scatter of 0.5 mag was
present in all of these relationships, and indicated that a second parameter
(beyond white dwarf mass) could be influencing the outburst luminosities of CNe.

Given the number of uncertainties that go into the derivation of the distances
using the secondary techniques, the reliability of these methods/relationships
has yet to be proven. What is needed to examine these techniques is high 
precision parallaxes. Using data obtained with the Fine Guidance Sensors on the 
$HST$, we have derived precise parallaxes for four CNe: V603 Aql, DQ 
Her, GK Per, and RR Pic. We use the distances for this small sample to explore
the nebular expansion parallax methods as applied to these sources, as well
as to test the various MMRD relations. In the next section we describe the
observations required to obtain parallaxes with the Fine Guidance Sensors,
in section 3 we provide a brief overview of how parallaxes are obtained
from such data, in section 4 we discuss the results for the individual
CNe, and in section 5 we state our conclusions.

\section{Observations}

The Fine Guidance Sensors (FGS), besides providing guiding for the other
science instruments on $HST$, can be used to obtain precision astrometry.
Details on the FGS instrument can be found in Nelan et al. (2011). The
main benefits of the FGSs are their large fields-of-view (3' x 10'), and high
dynamic range. Benedict et al. (2011) have thoroughly described how an
astrometric program is conducted with the FGS, and we refer the reader to
that discussion. Here we provide a brief overview of the process.

\subsection{$HST$ FGS Data}

A single ``POS Mode'' FGS observation consists of multiple measurements of 
the relative positions of the astrometric target and a set of reference frame 
stars. During this single $HST$ orbit, a typical astrometric sequence will 
result in the target being observed four or five times relative to the 
reference frame stars. The entire field is then observed at several well-separated
epochs. For a sufficiently bright target ($V$ $\leq$ 15.0) and a well-populated 
reference frame, ten orbits of FGS observations can produce parallaxes that 
have precisions of $\sigma_{\pi}$ $\leq$ $\pm$ 0.25 mas.  
For this particular program, with data from three $HST$ cycles (GO10912, 
GO11295, and GO11785), between eight and ten sets of astrometric data 
were acquired with {\it HST} FGS 1r for each CNe. Most of these data were 
obtained at epochs close to the time of maximum parallax factor (though 
occasionally tempered by two-gyro guiding constraints, see Benedict et al. 2010).
Thus, only small segments of the parallactic ellipses were observed for these 
targets. The various complete data aggregates span from 2.42 to 3.28 years.
 
Approximately forty minutes of spacecraft time were used to obtain each individual 
$HST$ data set. These data were then reduced and calibrated as detailed in 
McArthur et al.  (2001), and Benedict et al. (2002a, 2002b). At each epoch the 
positions of the reference stars and the target were measured several times to 
correct for intra-orbit drift (see Fig. 1 of Benedict et al. 2002a). Data were 
downloaded from the $HST$ archive and pipeline-processed. The FGS data
reduction pipeline extracts the measurements (the $x$ and $y$ positions from the 
fringe tracking, acquired at a 40 Hz rate, yielding hundreds of individual
measurements), extracts the median, corrects for the Optical Field Angle Distortion 
(c.f. McArthur et al. 2002), and adds the required time tags and parallax 
factors. 

\subsection{Ground-based Photometry and Spectroscopy}

As described below, to solve for the parallax of a program object using
the FGS, we need to estimate the parallaxes of the reference frame stars.
We use spectra to classify the temperature and luminosity class of each
star, and then combine these with $UBVRIJHK$ photometry to determine their
visual extinctions.

We obtained spectra of the reference frame stars for the three northern CNe 
using the Dual Imaging
Spectrograph\footnotemark[4]\footnotetext[4]{http://www.apo.nmsu.edu/arc35m/Instruments/DIS/} (``DIS'') on the 3.5 m telescope at the Apache Point Observatory. 
DIS simultaneously obtains spectra covering blue and red spectral regions,
and with the high resolution gratings (1,200 line/mm) provides dispersions
of 0.62 \AA/pix in the blue, and 0.58 \AA/pix in the red. For RR Pic,
we obtained spectra of the reference frame stars using the R$-$C
Spectrograph\footnotemark[5]\footnotetext[5]{http://www.ctio.noao.edu/spectrographs/4m\_R-C/4m\_R-C.html} on the Blanco 4 m telescope at Cerro Tololo Interamerican
Observatory (program 2009A-0009). The KPGL1 grating was used, and with the 
``Loral 3K'' detector, provided a dispersion of 1.01 \AA/pix.

Optical photometry for the fields of V603 Aql, DQ Her, and GK Per were
obtained using the robotic New Mexico State University (NMSU) 1 m telescope 
(Holtzman et al. 2010) at Apache Point Observatory. The NMSU 1 m is equipped 
with an E2V 2048 sq. CCD camera, and the standard Bessell $UBVRI$ filter set. 
Photometry of the field of RR Pic was obtained using the Tek2K CCD 
imager\footnotemark[6]\footnotetext[6]{http://www.ctio.noao.edu/noao/content/tek2k} 
on the SMARTS 0.9 m telescope at CTIO (program 2009A-0009). The images for
the four CNe fields, along with the appropriate calibration data, were obtained 
in the usual fashion, reduced using IRAF, and flux calibrated with observations 
of Landolt standards. 

Over the past decade, we have compiled an extensive set of template spectra 
covering a large range of temperature and luminosity classes in support
of our various FGS programs on both the APO 3.5 m,
and the Blanco 4 m. We perform MK classification of each of the reference
frame stars with respect to these templates, as well as use the temperature
and luminosity classification characteristics listed in Yamashita et al.
(1978). We find that for well exposed DIS spectra, our temperature
classifications are generally good to $\pm$ 1 subclass. For the lower
resolution CTIO data, however, there is more uncertainty, and we generally 
obtain spectral classifications with uncertainties of $\pm$ 2 subclasses.

Note that we are bound to encounter both subgiants and unresolved binaries
in a program with this many reference stars. For example, in Table 1 we 
identify DQ Her Ref \#01 as a potential binary because the parallax 
derived from our astrometric solution was much smaller than its spectroscopic
parallax. We identify possible binaries and subgiants from their large residuals
in the astrometric solution. First, the astrometric reference frame is modeled 
without the target CNe, as a check on the input spectroscopic parallaxes 
and proper motions. When the model fit to the reference frame is poor, we examine 
the reference stars individually, first by removal looking for a significant 
$\chi^{\rm 2}$ improvement, and secondly by treating those outliers as targets to 
redetermine a more likely a priori input parallax. We then confirm that this
re-classification is consistent with the spectroscopic and photometric data.
These redetermined spectroscopic parallax values are 
then used as input in the final astrometric model that includes the target CNe.

With the spectral classification of the reference stars complete, we then use the 
$UBVRI$ photometry we have obtained, in conjunction with $JHK$ photometry from 
2MASS, to derive the visual extinction to the sources using the reddening 
relationships
from Rieke \& Lebofsky (1985). Once determined, we can estimate spectroscopic
parallaxes using the absolute visual magnitude calibrations for main sequence
stars listed in Houk et al. (1997), and for giant stars using Cox (2000). We 
assemble all of the relevant data for the reference frame
stars in Table 1. The first column of this table lists the object 
identification, the second and third list the position (J2000), and the
fourth and fifth columns list the proper motions (in mas yr$^{\rm -1}$) as 
determined from our astrometric solution. The sixth column 
lists the derived spectral type of the reference frame star, the seventh column
is its $V$ magnitude, the eighth is its ($B - V$) color, the penultimate column
lists the visual extinction estimate, and the final column lists the parallax
(with error) computed from the astrometric solution (but advised by the
input spectroscopic parallax).

In Fig. 1, we plot the derived visual extinctions vs. the distances to
the reference frame stars listed in Table 1. In each figure we indicate the 
average line-of-sight value for the extinction in the direction of the program 
CNe using the IRSA
Galactic Dust Reddening and Extinction calculator\footnotemark[8]\footnotetext[8]{http://irsa.ipac.caltech.edu/applications/DUST/} (except 
for V603 Aql, where the line-of-sight extinction is enormous: A$_{\rm V}$ $>$
15 mag). In this figure we also denote the location of the program CNe
with crosses (discussed below).

\section{Deriving Parallaxes for the Program CNe}

With the $x$ and $y$ positions from the FGS 1r observations in hand, we proceed to
determine the scale, rotation, and offset ``plate constants'' for each epoch 
relative to a constraint epoch (the ``master plate"). We employ GaussFit 
(Jefferys et al. 1988) to simultaneously minimize the $\chi^2$ value for the 
following set of equations:

\begin{equation}
        x'  =  x + lc_x(\it B-V) 
\end{equation}
\begin{equation}
        y'  =  y + lc_y(\it B-V)
\end{equation}
\begin{equation}
\xi = Ax' + By' + C  - \mu_x \Delta t  - P_\alpha\pi_x
\end{equation}
\begin{equation}
\eta = Dx' + Ey' + F  - \mu_y \Delta t  - P_\delta\pi_y
\end{equation}

\begin{flushleft}
In the first two equations, $x$ and $y$ are the measured coordinates from the 
FGS, and $\it lc_x$ and $\it lc_y$ are the lateral color correction terms that
are dependent on the ($B - V$) color of each star. $A$, $B$, $D$, and $E$ are 
scale and rotation plate constants, $C$ and $F$ are offsets, $\mu_x$ and 
$\mu_y$ are the proper motions, $\Delta$t is the epoch difference from the 
mean epoch, $P_\alpha$ and $P_\delta$ are the parallax factors, while 
$\it \pi_x$ and $\it \pi_y$ are the parallaxes in $x$ and $y$. The 
parallax factors are obtained from a JPL Earth orbit predictor (Standish 1990), 
version DE405. This set of equations was used for deriving the parallaxes
of V603 Aql and RR Pic. For DQ Her, a four parameter solution was used (versus
the six parameter solution shown above), having identical scale factors 
for $x$ and $y$: $D$ $\equiv$ $-B$ and $E$ $\equiv$ $A$. For GK Per, we used
a similar scheme as that for DQ Her (identical scale coefficients in $x$ and $y$ ), 
but included additional radial scale terms into equations 3 and 4: 
\end{flushleft}

\begin{equation}
\xi = Ax' + By' + C + R_{x}(x^{2} + y^{2}) - \mu_x \Delta t  - P_\alpha\pi_x
\end{equation}
\begin{equation}
\eta = -Bx' + Ay' + F  + R_{y}(x^{2} + y^{2}) - \mu_y \Delta t  - P_\delta\pi_y
\end{equation}

\subsection{Input Modeling Constraints and Reference Frame Residuals} 

In our astrometric analysis, the reference star spectroscopic parallaxes and 
their proper motions from the PPMXL proper catalog (Roeser et al. 2010) are not
considered absolute, and were input as observations with associated errors. Typical
errors on the proper motions are of order 5 mas yr$^{-1}$ in each coordinate.
In addition, the lateral color and cross-filter calibrations, as well as the 
measured ($B - V$) color indices, were also considered as observations with error. 
Note that while the CNe exhibited orbitally
modulated brightness changes, their ($B - V$) colors remain relatively
constant over an orbit (see Bruch \& Engle 1994). Therefore we did not 
include in the modeling a time-dependent color correction value for any of 
the CNe. 

The calibration by McArthur et al. (2002) of the Optical Field Angle Distortion 
(OFAD) reduces the large distortions, of amplitude $\sim 1\arcsec$, seen across 
the field of the FGS 1r, to below 2 mas. The OFAD used for the present reduction and 
analysis of the FGS 1r data for the CNe has been updated with the post May 2009 
servicing mission observations (McArthur et al. 2012, in preparation). 
To determine if there might be systematic effects at the 1 mas level that could
be correctable, we investigated the reference frame $x$ and $y$ residuals against:
1) the position within the field-of-view, 2) the radial distance from the center of 
the field-of-view, 3) the $V$ magnitude and/or ($B - V$) color of the reference 
star, and 4) the epoch of observation. No such trends were detected. The final 
parallax and proper motion values (with errors)
obtained from our modeling of the FGS data for the program reference stars are 
listed in Table 1\footnotemark[9]\footnotetext[9]{A careful examination
of Table 1 will show that the number of reference stars used for the
astrometric solution for GK Per was smaller than for the other CNe. Three
of the program reference stars for this field (Ref \#3, \#6, and \#10)
showed large residuals that could not be reduced by multiple alternative models 
(e.g., models with different scale parameters, or models that omitted a priori values 
of parallax and proper motion for that reference star). One possible source
of such residuals is a field star located close to the target ($\leq$ 5", see
Nelan et al. 2011).  Another is that the object could be a binary star with a
significant reflex motion with an orbital period that is on order of the 
frequency of the observational epochs. Ref \#5 was dropped due to it being 
very faint ($V$ = 15.8) and red [($B - V$) = 1.1]. The good news is that
the remaining targets produced a very quiet reference frame.}

The casual reader might be surprised at the small size of the errors on the
parallaxes of the reference stars listed in Table 1. These small errors are 
informed by the input spectrophotometric parallaxes and their inherent error  
in a quasi-Bayesian manner. Because of the intrinsic width of the main sequence,
and the spectroscopic classification uncertainty, the reference 
star spectrocopic parallaxes typically have intrinsic input errors of order 
$\sim$ 25\%. Distant reference stars can have input spectroscopic parallaxes 
of order $\pi$ $\approx$ 0.25 mas, and thus the error bar on such parallaxes 
can be of order a 
few tens of $\mu$as. All errors in the reference star {\it a priori} data 
(proper motion and spectroscopic parallax inputs) are used, in a Bayesian 
fashion, by the GaussFit program to arrive at the final parameters for the 
reference frame. With eight to ten observational epochs, five or more 
reference frame stars per field, more than two of years of proper motion 
information, the final astrometric solution derives the reference frame 
parallaxes and errors. {\it The multiple measurements included into the 
astrometric analysis results in error bars on the parallaxes and proper motions 
that are smaller then their input values}. Note that the final reference 
star parallax errors are of order 8\%, as are the errors on our program 
objects (see below). The errors on the parallaxes and proper motions of the 
reference stars listed in Table 1 are uncorrelated. These errors are 
influenced by the quasi-Bayesian inputs, and thus are not truly independent 
measurements (in contrast to those of our program objects). The precision 
of the parallax for a program object is a direct consequence of the
quality of the astrometric solution for its reference frame. As 
demonstrated by Benedict et al. (2002b, their section 5.1), the error bars on 
the program object parallaxes derived using this methodology are conservative.

\subsection{The Parallaxes of V603 Aql, DQ Her, GK Per, and RR Pic}

For each of the CNe, we constrain $\pi_x = \pi_y$ in Equations 3 and 4 to
obtain the final parallaxes and proper motions listed in Table 2.
The precisions of the parallaxes in Table 2 are an indication of our internal, 
random error, and for the program CNe, these errors are $\approx$ $\pm$ 0.2 mas. 
To assess our external error, we have 
compared the parallaxes from previous FGS programs (Benedict et al. 2002b, 
Soderblom et al. 2005, McArthur et al. 2011) with results from {\it Hipparcos} 
(Perryman et al. 1997). Other than for the Pleiades (Soderblom et al. 2005), there 
are no significant differences between the results obtained with the FGS, and
with those from $Hipparcos$ for any object with high precision parallaxes.

Of the four program objects, the only one with a statistically significant
$Hipparcos$ parallax is V603 Aql. Due to its faintness ($V$ = 11.7, Bruch
\& Engel 1994), V603 Aql was a difficult target for $Hipparcos$. The
original $Hipparcos$ catalog lists $\pi_{abs}$ = 4.21 $\pm$ 2.59 mas. The van 
Leeuwen (2007) re-reduction of the $Hipparcos$ data yielded $\pi_{abs}$ = 
4.96 $\pm$ 2.45 mas for V603 Aql. Both determinations agree with our measurement
($\pi_{abs}$ = 4.011 $\pm$ 0.137 mas), given their significant error bars.

\subsection{The Lutz-Kelker-Hanson Correction to M$_{\rm V}$}

As noted long ago by Trumpler \& Weaver (1953), a systematic error is introduced
into the calibration of the luminosities for a group of objects when using parallax. 
Due to the fact that in nearly every stellar population, the number of stars in a 
sample increases with distance, stars with overestimated parallaxes will 
outnumber those with underestimated parallaxes. Lutz \& Kelker (1973) showed
that the size of the bias depends only on the ratio of $\sigma_{\pi}$/$\pi$. 
Here we have used the general formulation of Hanson (1979) to determine 
the corrections for the program CNe. We calculate the Lutz-Kelker-Hanson (``LKH'') 
bias for our CNe presuming that they all belong to the same class of object (old 
disk stars), and report the LKH correction to be applied to the object's absolute
visual magnitude in the final column of Table 2. Given the uncertainties
in the peak visual magnitudes of the program CNe, these small adjustments
are unimportant in characterizing the outbursts of the program CNe, and
will be ignored in what follows.

\section{Results}

With the astrometric results, we investigate the outbursts of the program CNe with 
respect to their light curve decline rates. Below we assemble both the published
$t_{\rm 2}$ and $t_{\rm 3}$ decline rates for the program novae, as well
as review their light curves to examine the long-established values for their
maximum visual magnitudes. The MMRD relationships critically depend on
having precise values for both of these quantities, thus we feel it is important
to review the origins of the previously published values for those data. Having 
precisely known distances also 
allows for the investigation of the expansion of the nebulae produced in each of 
the outbursts. Downes \& Duerbeck (2000) provide a summary of the outbursts of
each of these CNe, including distance estimates derived using
nebular expansion parallaxes. We compare the new astrometric distances 
with the distances from the nebular expansion parallaxes in Table 3.
Except for GK Per, the astrometric distances for the
CNe turn out to be smaller than those estimated by Downes \& Duerbeck.
We order our discussion alphabetically by constellation name.

\subsection{V603 Aquilae}

V603 Aql erupted in 1918 June, and due to its brightness, a comprehensive
light curve was compiled (Campbell 1919). There were also numerous spectroscopic
observations of the outburst, and those data have been discussed by Wyse (1940).
The light curve presented by Campbell shows that the nova reached m$_{\rm v}$ 
= $-$1.1 on 9 June 1918. In Fig. 2, we present the light curve of V603 Aql 
close to this date from the data in Table III of Campbell. Note that there
are nine visual magnitude estimates that have the nova as being brighter than
m$_{\rm v}$ = $-$1.1. Six of these are due to E. E. Barnard (Yerkes). Note
that we have used the ``Corrected Magnitudes", for which Campbell accounted
for the ``bias'' of the observer. In fact, Barnard reported that the nova
peaked at m$_{\rm v}$ = $-$1.5 on JD24211754.94, to which Campbell subsequently
applied a correction of +0.1 mag. One might discount Barnard's observations
given the difficulty of estimating the magnitude of something that was so
much brighter than any naked eye stars of that season, but on JD24211754.78, W. H. 
Pickering (Harvard College Observatory, Mandeville, Jamaica)
estimated m$_{\rm v}$ = $-$1.2; fifteen minutes later (JD24211754.79) Barnard 
derived the same brightness.

The consistency of the data, and the reputation of the observers in question,
suggests that V603 Aql easily exceeded the commonly quoted value of m$_{\rm v}$ = 
$-$1.1 at visual maximum. These data support a value of {\it at least} 
m$_{\rm v}$ = $-$1.4 for its maximum. The discrepant data point near those 
of peak brightness, m$_{\rm v}$ = $-$0.7 (at JD24211754.98), is due to Conroy 
(1918), an amateur astronomer based in Los Angeles. Conroy indicates that at the 
time of his estimate, V603 Aql was ``much bluer than Vega'', suggesting that it 
had not yet reached visual maximum.
It is interesting to note in his spectroscopic survey of ``old novae'' Humason 
(1938) lists m$_{\rm v}$ = $-$1.4 for the maximum of V603 Aql.
We tabulate the outburst characteristics of V603 Aql, and the other program novae,
in Table 4. 

As discussed above, it has long been suggested that there is a relationship for
CNe between their absolute visual magnitudes at maximum, and the rate of 
decline in their light curves from visual maximum. We have averaged the reported 
$t_{\rm 2}$ and $t_{\rm 3}$ values from the literature (Duerbeck 1987; 
McLaughlin 1939; and Strope et al. 2010) for V603 Aql to arrive at the values 
listed in Table 4. 
The published values of these two quantities are all quite similar, 
due to the rather smooth decline of the light curve from maximum. Note that if 
we assume V603 Aql actually reached m$_{\rm v}$ = $-$1.4 at peak, the resultant 
$t_{\rm 2}$ and $t_{\rm 3}$ values are reduced to 1.5 d, and 6 d, respectively.
The extinction to V603 Aql is low, with E($B - V$) = 0.07 (Gallagher \&
Holm 1974). Using this, the new parallax, and m$_{\rm v_{max}}$ = $-$1.4, we derive 
an absolute visual magnitude at maximum of M$_{\rm V_{max}}$ = $-$8.60. 
Given the estimate for the mass of its white dwarf, M$_{\rm 1}$ = 1.2 $\pm$ 0.2
(Arenas et al. 2000), at its peak, V603 Aql exceeded the Eddington limit by
$\sim$ 1.7 mag, but its super-Eddington phase (at visual wavelengths) only lasted 
$\approx$ 48 hr.

Besides an extensive discussion of the spectra of V603 Aql, Wyse (1940)
compiled measurements of the size of the expanding nebular shell from the
eruption of V603 Aql first noticed by Barnard (1919). With a precise parallax,
we can determine the expansion velocity required to reproduce the observations.
We plot the angular measurements of the disk of V603 Aql versus the time
since outburst in Fig. 3. It is apparent from this figure that the expansion
velocity needed to produce an ejected shell that evolved in the observed way was
$\approx$ 1,100 km s$^{\rm -1}$. This value is much lower than the published 
velocities
of the ``principal absorption'' components of 1,500 km s$^{\rm -1}$ (McLaughlin
1940) or 1,700 km s$^{\rm -1}$ (Payne-Gaposchkin 1957). Since it has long
been believed that the principal absorption component is the velocity of the
bulk of the ejecta (e.g., Payne-Gaposchkin), it is somewhat surprising that the 
observed expansion of the nebula indicates a much lower velocity.

A possible way to reconcile these observations comes from a model of the ejected
shell of V603 Aql constructed by Weaver (1974). Weaver finds that the spectroscopic 
record is consistent with a shell that has its long axis pointed towards the Sun.
A recent estimate of the orbital inclination of the underlying binary 
arrives at $i$ = 13$^{\circ}$ (Arenas et al. 2000). Thus, we view V603 Aql
nearly pole-on. If we ratio the values of the observed ``equatorial'' expansion 
velocity with the principal absorption velocities, we derive an ellipsoid that
has a ratio of its minor to major axes of 0.65 $\leq$ b/a $\leq$ 0.73. This
is similar to that of DQ Her (see below), suggesting that interaction with
the accretion disk and/or secondary star acts to slow the progress of the
ejecta in the plane of the binary star system.

\subsection{DQ Herculis}

DQ Her erupted in 1934, reaching maximum on 22 December. Monographs by
McLaughlin (1937) and Beer (1974) thoroughly discuss the spectroscopy of the
outburst of this prototypical dust-producing nova. DQ Her is classified as
a moderate speed nova, and we tabulate the means of the decline rates
taken from the literature (McLaughlin 1939, Strope et al. 2010, Duerbeck 1987)
in Table 4. In addition to those published values, we have examined the light 
curve data published by Gaposchkin (1956) and the light curve assembled by
Beer (1974), to derive additional values of $t_{\rm 2}$ = 80.4 d, 67 d, and
$t_{\rm 3}$ = 94.3 d, 94 d, respectively, and these data been incorporated into 
the means listed in the Table 4.
Downes \& Duerbeck (2000) list $t_{\rm 2}$ = 39 d for DQ Her, but this
value is due to a short-lived dip at the end of 1935 March, from which the nova 
recovered, after which it resumed the more general decline rate that was present 
before this event. We have not incorporated that value into the $t_{\rm 2}$
mean for DQ Her.

The published data for the light curve maxima are all quite similar and
lead to the mean of v$_{\rm max}$ = 1.3 listed in Table 4. The value of the
visual extinction to DQ Her is somewhat more uncertain. The commonly quoted
value is A$_{\rm V}$ = 0.35, but this appears to be due to the value quoted
in Ferland et al. (1984).  Ferland et al. state that this value is the 
line-of-sight extinction for galaxies in this direction from de Vaucouleurs 
et al. (1976). Analysis of $IUE$ spectra of DQ Her by Verbunt (1987) gives a
similar value for the extinction: E($B - V$) = 0.1. As shown in Fig. 1, 
if we use this value, DQ Her has a significantly higher extinction than its
reference frame stars. The IRSA data base gives a much lower value of A$_{\rm V}$ 
= 0.13 for the line-of-sight extinction at the location of DQ Her. This latter 
estimate is perfectly consistent with the values we derived for the astrometric 
reference frame stars. We find that DQ Her appears to suffer from an anomalously 
high extinction. The most likely explanation is excess {\it local} extinction from 
circumstellar material, perhaps from the dust shell created in the eruption. Note 
that Evans (1991) detected molecular gas around this object, and DQ Her was also
detected at both 60 and 100 $\mu$m by $IRAS$ (Harrison \& Gehrz 1988, Callus et 
al. 1987, Dinerstein 1986).

To determine the absolute visual magnitude of DQ Her at outburst maximum, we have
incorporated the value of A$_{\rm V}$ = 0.31 from Verbunt (1987). With $d$ =
386 pc and  v$_{\rm max}$ = 1.3, this leads to M$_{\rm V_{\rm max}}$ = $-$6.94. If 
the excess extinction is due to the dust shell created in April of 1935, however, 
DQ was slightly fainter (0.18 mag) at visual maximum: M$_{\rm V_{\rm max}}$ = 
$-$6.76. This shows that at its peak, assuming M$_{\rm 1}$ = 0.60 $\pm$ 0.07 
M$_{\sun}$ (Horne et al. 1993), the luminosity of DQ Her exceeded the Eddington limit
by about 0.7 mag.

Vaytet et al. (2007) provide the most recent analysis of the size and structure
of DQ Her's ellipsoidal ejected shell, including the detection of 
clumps/knots that appear to be ablated by a strong stellar wind aligned with the 
poles of the binary. They find the radial distances to the center of the ring of
the ejected shell in the major and minor axis directions to be $a$ = 25.31 $\pm$ 
0.44, and $b$ = 18.70 $\pm$ 0.44 arcseconds, respectively (epoch 1997.82).
They derived a maximum line-of-sight velocity of 370 km s$^{\rm -1}$, from
which they calculated a distance of 525 pc $\pm$ 28 pc. This number is
substantially larger than our astrometric value.

If we use the new parallax and the Vaytet et al. measurements, we derive expansion 
velocities of 368 and 272 km s$^{\rm -1}$ along the major and minor axes, 
respectively. The mean of these two velocities (320 km s$^{\rm -1}$) is very 
close to the value of the velocity of the principal absorption component listed 
by McLaughlin (1940): 315 km s$^{\rm -1}$. Ferland (1980) quote that analysis of 
the emission lines from the nebular shell gave a velocity of 320 $\pm$ 20 km 
s$^{\rm -1}$. While the average of the velocities of the two shell axes is 
consistent with the assumption that the principal velocity component observed near 
maximum light is associated with the bulk of the ejecta for DQ Her, the details are 
not. 

We plot the values for the expanding shell of DQ Her in Fig. 4. The early 
micrometer measures by Kuiper (1941) are only of the major axis. The first 
measure of the minor axis dimension is due to Baade (1940). Duerbeck (1987)
tabulates the measures up to 1984. The last three measurements plotted in Fig. 4
are due to Slavin et al. (1995), Vaytet et al. (2007), and our own measurement of
an unpublished $HST$ WFPC2 image ($HST$ Proposal ID: 6060) obtained on 1995 
September 4. As shown in Fig. 4, the diameter of the major axis of the nebula
suggests an expansion velocity in excess of 320 km s$^{\rm -1}$, while the minor 
axis of the nebula is much smaller than would be expected if it was expanding
at this rate. In fact, these data are suggestive of a slowing in the 
rate of expansion of the minor axis that appears to have started around 1970, when 
the shell's distance from the central binary was $\sim$ 1.7 $\times$ 10$^{\rm 16}$ cm. 
This suggests to us that there is pre-existing material 
into which the ejected shell has collided that has acted to retard its progress. 
Perhaps this is evidence for a circumbinary disk of material (c.f., Dubus et al. 2002).
DQ Her is an eclipsing binary, and we view the system almost edge on: $i$ = 
86.$^{\circ}$5 (Horne et al.  1993), and thus circumbinary material along the
minor axis would be aligned with the plane of the underlying binary. Such a structure 
could also be responsible for the 
excess extinction, the $IRAS$ detections, and the H$_{\rm 2}$ emission.

\subsection{GK Persei}

The first bright nova of the twentieth century was discovered on 1901 February
21 by T. D. Anderson. Campbell (1903) compiled the light curve data for GK Per, and
concluded that it reached a visual maximum of m$_{\rm v_{\rm max}}$ = 0.2 on 
February 23$^{\rm rd}$. However, there are six estimates in Campbell's Table 
II that are brighter than his quoted peak for this nova. It is unclear why those 
data were ignored, as they come from respected observers: E. C. Pickering, W. H. 
Pickering, and A. J.  Cannon. In the cases of these three observers, they all
quoted GK Per as being ``two grades'' (0.2 mag) brighter than Capella ($V$ =
0.08). Fortunately, for much of the data set, Campbell lists the actual brightness
estimates relative to various comparison stars, and we can use modern values for 
the $V$ magnitudes of the comparison stars to regenerate the light curve of GK 
Per. We plot these ``calibrated'' magnitudes for GK Per in Fig. 5 as solid circles.
If the comparison stars were not listed, the magnitudes in Campbell's Table II
were used and are plotted as crosses in Fig. 5. Clearly, GK Per was at least as 
bright as $V$ = 0.0 at maximum. This is the value
quoted by Humason (1938) in his tabulation of novae maxima, and is what we
have listed in Table 4. Additional support for this result is found in the
popular literature of the time: ``On February 22, 1901, a marvelous new star was 
discovered by Doctor Anderson of Edinburgh, not very far from Algol. No star had 
been visible at that point before. Within twenty-four hours the stranger had 
become so bright that it outshone Capella. In a week or two it had visibly faded, 
and in the course of a few months it was hardly discernible with the naked eye.''
(G. P. Serviss, as quoted by Lovecraft, 1919).

The initial decline of GK Per from maximum was very smooth, and the mean values
for $t_{\rm 2}$ and $t_{\rm 3}$ have small error bars. The extinction to
GK Per, however, is quite large: A$_{\rm V}$ = 0.96 mag (averaging the
values from Wu et al. 1989, and Shara et al. 2012). With $d$ = 477 pc, this
leads to M$_{\rm V_{\rm max}}$ = $-$9.35. As noted above, it is quite possible
that GK Per was 0.1 mag more luminous than this at the time of visual maximum.

The structure of the shell of GK Per has been extensively investigated by 
Seaquist et al. (1989) and Shara et al. (2012). Slavin et al. (1995) reported 
that the shell had
dimensions of 103'' $\times$  90'' on 1993 September 10. With the parallax,
those dimensions correspond to expansion velocities of 1,256 and 1,100 km 
s$^{\rm -1}$ for the major and minor axes, respectively. McLaughlin (1940)
lists the principal absorption component of GK Per having a velocity of
$v_{\rm principal}$ = 1,300 km s$^{\rm -1}$, similar to that derived for
the major axis. The nebula ejected by GK Per is asymmetric and Seaquist et al.
discuss a scenario where the nova erupted within a circumstellar cloud that is
several parsecs across, with which it is now interacting. They propose it is this
material that was responsible for the light echos observed following 
outburst (c.f., Perrine 1902).

GK Per is an unusual CNe, having the second longest orbital period
known: 1.9968 d (Morales-Reuda et al. 2002). Clearly, the secondary star must be 
substantially larger than a main sequence star to fill its Roche lobe and 
transfer matter to the white dwarf primary. We can use the new distance 
and the implied Roche lobe geometry to investigate the nature of this 
system. Sherrington \& Jameson (1983) list GK Per as having $K$ = 10.14 
at minimum light. As shown in Harrison et al. (2007), the K2 secondary star 
of GK Per completely dominates the spectral energy distribution in the 
near-infrared. This leads to M$_{\rm K}$ = 1.65 for the subgiant 
secondary star. A K2V has M$_{\rm K}$ = 4.15, thus the secondary star 
in GK Per is exactly ten times more luminous than its main sequence 
counterpart. If we compare the secondary star of GK Per to the K2 dwarf 
$\epsilon$ Eridani (R = 0.74 $\pm$ 0.01 R$_{\sun}$, Baines \& 
Armstrong 2012), we calculate that its mean radius is R$_{\rm 2}$ = 
1.63 $\times$ 10$^{\rm 11}$ cm. Using the relationships in Warner (1995) 
between the orbital period, semi-major axis, the mass ratio ($q$ = 0.55 
$\pm$ 0.21; Morales-Reuda et al. 2002), and the Roche lobe radius of 
the secondary star, we derive that the white dwarf in GK Per has a 
mass of M$_{\rm 1}$ = 0.77$^{\rm +0.52}_{\rm -0.24}$ M$_{\sun}$ (where 
the limits on the mass only contain the errors associated with $q$). 
This simple calculation shows that the mass of the white dwarf in GK Per 
is not unusual when compared to other CVs (see Cropper et al. 1998), 
though a new study to refine the value of the mass ratio is clearly 
warranted.
If we assume that the bolometric correction at visual maximum is zero, then
for the derived white dwarf mass, GK Per exceeded the Eddington luminosity by a 
factor of fourteen at its peak, and remained above this limit for at least 10 days.

\subsection{RR Pictoris}

RR Pic was a very slow nova that erupted in 1925. A light curve of its
outburst can be found in Spencer Jones (1931), in which visual maximum occurs
on 7 June, 1925. We have compiled the $t_{\rm 2}$ and $t_{\rm 3}$ 
decline rates from the literature (Downes \& Duerbeck 2000; McLaughlin 1939; 
Duerbeck 1987; and Strope et al. 2010), and to those we add the values of 
$t_{\rm 2}$ = 82 d and $t_{\rm 3}$ = 122 d from our analysis of the light
curve compiled by Campbell (1929), to construct the mean values listed in
Table 4. Spencer 
Jones notes the unusual behavior of this object, in that it was later found on 
patrol photographs to be at m$_{\rm v}$ = 3.0 some six weeks prior to discovery.
Spencer Jones also notes that an amateur stated that he was confident that
no new naked eye stars were present at this position only four days prior to 
discovery. Given that there was a two month gap between the patrol photographs
showing it to clearly be at minimum (18 February), and the pre-discovery 
observation, Spencer Jones suggests that perhaps the true visual maximum of
this object was missed, and the maximum that occurred in June of that year
was a secondary event.

The line-of-sight extinction to RR Pic is low, with the mean of the published
values (Verbunt 1987; Krautter et al. 1981; Williams \& Gallagher 1979) giving
A$_{\rm V}$ = 0.13 mag. The mean value of the visual maximum from the
light curve sources listed above is V$_{\rm max}$ = 1.1 $\pm$ 0.1; this
leads to M$_{\rm V_{max}}$ = $-$7.61, about 1 mag above the Eddington limit
for a 1 M$_{\sun}$ white dwarf. This luminosity is larger than
expected given that RR Pic was a slower nova than DQ Her. If maximum
absolute visual magnitude is assumed to be directly correlated with the light 
curve decline rate, RR Pic should have been {\it less} luminous than DQ Her since
it was the slower nova.
Instead, RR Pic was almost twice as luminous as DQ Her at their respective
peaks. This may be additional evidence that the true visual maximum of 
this CNe was missed.

McLaughlin (1940) lists the principal component expansion velocity as 
285 km s$^{\rm -1}$, while Payne-Gaposchkin (1957) has $v_{\rm principal}$ = 
310 km s$^{\rm -1}$. Both Williams \& Gallagher (1979) and Gill \& O'Brien (1998) 
present analyses of the nebular shell of RR Pic. At the epochs of those
two observations, a freely expanding spherical shell with $v$ = 310 km s$^{\rm -1}$
would have diameters of 13.0" $\pm$ 1.2" and 17.6" $\pm$ 1.7", respectively. The actual 
shells had dimensions of 23" $\times$ 18" (Williams \& Gallagher 1979) and 
30" $\times$ 21" (Gill \& O'Brien 1998). The observed shells are significantly larger
than would be expected if the velocity of the principal absorption 
component measured near the June maximum was correct. Spencer Jones
lists a variety of other velocity systems for RR Pic, but none of them appear
to correspond to the velocity ($\sim$ 430 km s$^{\rm -1}$) required to create
the observed shell sizes.

RR Pic was observed with WFPC2 on the $HST$ in 1999 (26 February, Prop. ID \#6770).
We have analyzed those data and find that the {\it centers} of the main shell 
features are separated by 22.7". These features are quite diffuse, but appear
correspond to the ``equatorial ring'' condensations visible in the images 
presented by Gill \& O'Brien (1998). To be of this size requires an 
expansion velocity of 380 km s$^{\rm -1}$. While this is closer to the observed
principal velocity than implied by the previous studies, its remains 20\% larger.
Note that even the first visual micrometer observations of the young shell
(van den Bos \& Finsen 1931) are consistent with this higher than expected ejecta
velocity. Unlike the results for the previous objects, it is not as obvious
why the nebular expansion parallax method fails for RR Pic.

\section{Discussion}

To fully understand the outbursts of CNe, we need to have precise distances to 
accurately calorimeter their outbursts, and to allow us to examine the shell
ejection process. Theory suggests that fast novae occur on massive white
dwarfs, have the most luminous outbursts, have light curves exhibit the most rapid 
decline rates, and their ejecta have the highest expansion velocities. While our 
sample is tiny, having precise parallaxes for four objects sheds new light 
into the difficulties of making broad assumptions about the behavior of CNe.

In Fig. 6 we have plotted the absolute magnitudes at visual maximum 
(M$_{\rm V_{\rm max}}$) 
versus the {\it log} of their light curve decline rates (both $t_{\rm 2}$
and $t_{\rm 3}$) for the four program CNe. We also plot the various 
MMRD relationships discussed in Downes \& Duerbeck (2000). Both GK Per
and DQ Her fall very close to the linear relationships for $t_{\rm 2}$
and $t_{\rm 3}$. While the older ``arctangent'' law of Della Valle \& Livio
(1995) works for both V603 Aql and DQ Her. RR Pic remains an outlier in
all cases. As Downes \& Duerbeck show, there remains a scatter of $\sim$ 0.5
mag around the various relationships, and it was hoped that those inaccuracies 
were due to flaws in the secondary distance estimation techniques. The 
astrometric results show that such discrepancies remain.

We believe, however, that there are possible (partial) explanations for why both 
RR Pic and V603 Aql are so discrepant. For RR Pic, there appears to be sufficient
evidence to suggest that the 1925 June 6 maximum was a secondary event.
If one presumes the initial maximum reached to the same level as the 
June maximum (m$_{\rm v}$ = 1.1), then if it erupted sometime after 1925 February 18 (the last
quiescent patrol photograph), and was third magnitude on 1925 April 13,
then it would have had $t_{\rm 2}$ $\leq$ 54 d. This moves it much closer
to the linear relationship for $t_{\rm 2}$ (it needs to have $t_{\rm 2}$ = 29 d
to fall on exactly the line). There have been a number of CNe that have been 
observed to have complex light curves similar to that needed to have been 
exhibited by RR Pic to reconcile its decline rate with its observed absolute 
magnitude (see the ``C-class'', and ``J-class'' CNe light curves in Strope et al. 
2010). The main difficulty with this scenario is that it is hard to believe that 
a first magnitude nova would have escaped detection, given that it was
reasonably well placed for evening viewing in March and April.

While V603 Aql does fall near the older arctangent law, its outburst was
quite underluminous for the speed class when compared to the two linear laws. We 
argued above that the data suggests V603 Aql exceeded the commonly tabulated value 
of m$_{\rm V_{\rm max}}$ = $-$1.1, probably reaching m$_{\rm V_{\rm max}}$ =
$-$1.4.  The question was whether it was even brighter than this. To get 
the absolute visual magnitude to fall closer to the linear law lines, V603 Aql 
would have had to have reached m$_{\rm V_{\rm max}}$ $\approx$ $-$2.4. There is
a seven hour gap in the light curve right at the time of visual maximum,
so it is quite possible that the true maximum was missed. But an extrapolation
of the rise to the observed maximum suggests that it would have required a
sudden change in slope to exceed m$_{\rm V_{\rm max}}$ $\approx$ $-$1.8.
Thus, there truly appears to be a difference between the absolute visual
magnitudes at maximum of V603 Aql and GK Per, even though their light curve
decay rates were quite similar. Given that the principal
ejecta velocities were higher for V603 Aql, suggests that it had the more
violent eruption. This implies that there must be another 
parameter besides white dwarf mass that acts to govern the luminosity of CNe 
outbursts. The fact that the nebular remnant of GK Per is still visible more
than 100 yr after outburst, while that of the more recent V603 Aql is not, 
indicates that the shells ejected by these two objects were quite different.

Downes \& Duerbeck (2000) also explore the suggestion (originally due to 
Buscombe \& de Vaucouleurs 1955) that all CNe have
the same absolute visual magnitudes 15 days after visual maximum, finding
 M$_{\rm V_{\rm t=15d}}$ = $-$6.04. Using the published light curves we find 
that there is more than
a 1.5 mag spread between V603 Aql (M$_{\rm V_{\rm t=15d}}$ = $-$4.4) and
GK Per (M$_{\rm V_{\rm t=15d}}$ = $-$6.0) at this time in their outbursts. We 
conclude that using the 
decline rates of CNe light curves to obtain reliable distances is not possible. 
The fastest, and therefore most luminous novae, need near constant photometric 
monitoring to fully constrain their peak brightnesses. Even then, there are 
intrinsic differences in their outbursts that limits their 
value as standard candles.

While CNe may not be the best objects to use for extragalactic distance
estimates, the question is whether we can actually determine the distances to
individual CNe to attempt to characterize their outbursts. The most reliable
secondary distance estimation technique we have is the nebular expansion 
parallax method. This technique remains the main source of distances to CNe, and
has been used to calibrate the various MMRD relationships. The news on this
front is also not very heartening. For V603 Aql, we found that the velocity of the
principal absorption component (1,500 $\leq$ v$_{\rm principal}$ $\leq$ 1,700
km s$^{\rm -1}$) was much higher than the observed expansion rate of the
ejecta: 1,100
km s$^{\rm -1}$. For DQ Her, the results were slightly better, except it
appears that the expansion rate of the ejecta in the plane of the binary
(the minor axis of the nebula) has slowed over the last 40 years. For RR
Pic, the nebula is expanding much more rapidly than than the derived 
v$_{\rm principal}$. Only for GK Per is the expansion parallax in accordance
with expectations.

The good news is that we believe we can resolve the discrepancies for
the three discordant CNe. The smaller than expected expansion of the shells
of DQ Her and V603 Aql appears to be due to the interaction of the ejecta with 
the secondary star or with material that lies within the plane of the 
underlying binary star orbit. Note that we calculated that
V603 Aql and DQ Her appear to have similar ratios of the minor to major axes
for their ellipsoidal shells. Unfortunately, to determine this requires
one to construct a model for each of these shells, stressing the importance
of high resolution spectroscopy throughout the outburst and decline of CNe,
as well as follow-up, multi-epoch imaging. For 
RR Pic, the discrepancy can be eliminated if we assume that the observed maximum
was in fact a secondary maximum. This simply requires a slightly higher 
principal ejecta velocity at the time of its true maximum. 

Lloyd et al. (1997) have simulated the effect that the underlying binary has
in shaping the shells of CNe. The results from Lloyd et al. suggest that the
shells of fast novae should mostly ignore the underlying binary. But this
is not the case for V603 Aql. It appears that the shell of V603 Aql
was as non-spherical
as that of the much slower DQ Her. All three of the CNe for which the
nebular expansion parallax technique does not work have much shorter orbital 
periods (P$_{\rm DQ}$ = 4.64 hr, P$_{\rm RR}$ = 3.48 hr, P$_{\rm V603}$ = 
3.32 hr) than for the concordant GK Per (P$_{\rm GK}$ = 48.1 hr). This suggests to 
us that the interaction of the secondary star with the ejecta is probably
more important than the simulations indicate.

One of the unfortunate aspects of the current CNe sample is that three
of the objects have been classified as ``Intermediate Polars'', CVs that are
believed to have highly magnetic white dwarf primaries (B $\lessapprox$ 1 MG).
Such objects are identified by having coherent periodicities that are shorter
than their orbital periods, assumed to originate from processes occuring at the
magnetic poles of the rapidly rotating white dwarfs in these systems. It is unclear
if strong magnetic fields play any role in shaping the outburst or the
ejecta of CNe (Livio et al. 1988, Nikitin et al. 2000). The fact that RR Pic is not 
an Intermediate Polar (Pek\"{o}n \& Balman 2008), but also has a discrepant nebular
expansion parallax, indicates that the presence of a strong magnetic field does not 
appear to dramatically affect the shell ejection process.

The luminosities of the outbursts of CNe are obviously more complex than being a 
simple function of the mass of the white dwarf primary in the underlying binary. 
It would be extremely useful to have additional
parallaxes to construct a larger sample of objects with precise distances
but, unfortunately, few CNe have minimum magnitudes that will allow for 
precise parallaxes even with the $GAIA$ mission. The subset of those 
that had outbursts with the quality of data necessary to deconvolve the 
nature of their outbursts is even smaller. Thus, progress on characterizing CNe 
outbursts will be better served by observations of future CNe. This will require 
more thorough all-sky monitoring to insure that the light curves of 
these objects have better temporal coverage. In addition, however, multi-epoch 
interferometric, high resolution imaging, and moderate resolution spectroscopic 
observations of these CNe will also be required. 

\acknowledgements Support for program AR12617 was provided by NASA through a grant 
from the Space Telescope Science Institute, which is operated by the Association of 
Universities for Research in Astronomy, Inc., under NASA contract NAS 5-26555.

\begin{flushleft}
{\bf References}\\
Arenas, J., Catal$\acute{a}$n, M. S., Augusteijn, T., \& Retter, A. 2000, MNRAS, 
311, 135\\
Baade, W. 1940, PASP, 52, 386\\
Baines, E. K., \& Armstrong,, J. T., 2012, ApJ, 744, 138\\
Barnard, E. E. 1919, ApJ, 49, 199\\
Beer, A. 1974, Vistas in Astronomy, 16, 179\\
Benedict, G. F., et al. 2011, AJ, 142, 187\\
Benedict, G. F., McArthur, B. E., Bean, J. L., Barnes, R., Harrison, T. E.,
Hatzes, A., Martioli, E., \& Nelan, E. P. 2010, AJ, 139, 1844\\
Benedict, G. F., McArthur, B. E., Feast, M. W., Barnes, R., Harrison, T. E.,
Patterson, R. J., Menzies, J. W., Bean, J. L., \& Freedman, W. L. 2007,
AJ, 133, 1810\\
Benedict, G. F. et al. 2002b, AJ, 124, 1695\\
Benedict, G. F. et al. 2002a, AJ, 123, 473\\
Bruch, A., \& Engel, A. 1994, A\&AS, 104, 79\\
Buscombe, W., \& de Vaucouleurs, G. 1955, Observatory, 75, 170\\
Callus, C. M., Evans, A., Albinson, J. S., Mitchell, R. M., Bode, M. F., 
Jameson, R. F., King, A. R., \& Sherrington, M. 1987, MNRAS, 229, 539\\
Campbell, L. 1929, Harvard Bull., 835\\
Campbell, L. 1919, Harvard Ann., 81, 113\\
Campbell, L. 1903, Harvard Ann., 48, 39\\
Cannizzo, J. K., Still, M. D., Howell, S. B., Wood, M. A., \&
Smale, A. P. 2010, ApJ, 725, 1393\\
Conroy, C. C. 1918, MNRAS, 79, 37\\
Cox, A. N. 2000, ``Allen's Astrophysical Quantities'', (AIP Press: New York)\\
Cropper, M., Ramsay, G., \& Wu, K. 1998, MNRAS, 293, 222\\
de Vaucouleurs, G., de Vaucouleurs, A., \& Corwin, J. R. 1976, in {\it
Second reference catalogue of bright galaxies}, (Austin: University of Texas Press)\\
Della Valle, M., \& Gilmozzi, R. 2002, 296, 1275\\
Della Valle, M., \& Livio, M. 1995, ApJ, 452, 704\\
Dinerstein, H. L. 1986, AJ, 92, 1381\\
Downes, R. A., \& Duerbeck, H. W. 2000, AJ, 120, 2007\\
Ferland, G. J., Williams, R. E., Lambert, D. L., Shields, G. A., Slovak, M.,
\& Gondhalekar, P. M. 1984, ApJ, 281, 194\\
Ferland, G. J., 1980, The Observatory, 100, 166\\
Gill, C. D., \& O'Brien, T. J. 1998, MNRAS, 300, 221\\
Hachisu, I., \& Kato, M. 2010, ApJ, 709, 680\\
Hanson, R. B. 1979, MNRAS, 186, 875\\
Harrison, T. E., Johnson, J. J., McArthur, B. E., Benedict, G. F., Szkody,
P., Howell, S. B., \& Gelino, D. M. 2004, 127, 460\\
Harrison, T. E., \& Gehrz, R. D. 1988, AJ, 96, 1001\\
Holtzman, J. A., Harrison, T. E., Coughlin, J. L. 2010, Adv. Astron.,
193086\\
Horne, K., Welsh, W. F., \& Wade, R. A. 1993, ApJ, 410, 357\\
Houk, N., Swift, C. M., Murray, C. A., Penston, M. J., \& Binney, J. J. 1997, in
Proc. ESA Symposium on Hipparcos$—$Venice 1997 (ESA SP-402; Noordwijk:
ESA), 279 \\
Humason, M. L. 1938, ApJ, 88, 228\\
Jefferys, W. H., Fitzpatrick, M. J., \& McArthur, B. E. 1988, Celest.
Mech., 41, 39\\
Krautter, J., Vogt, N., Lare, G., Wolf, B., Duerbeck, H. W., Rahe, J., \& 
Wargau, W. 1981, A\&A, 102, 337\\
Kuiper, G. P. 1941, PASP, 53, 330\\
Livio, M., Shankar, A., \& Truran, J. W. 1988, ApJ, 330, 264\\
Lloyd, H. M., O'Brien, T. J., \& Bode, M. F. 1997, MNRAS, 284, 137\\
Lovecraft, H. P. 1919, in ``Beyond the Wall of Sleep'', {\it Pine Cones}, 1, 2\\
Lutz, T. E., \& Kelker, D. H. 1973, PASP, 85, 573\\
McArthur, B. E., et al. 2002, in ``The 2002 HST Calibration Workshop: Hubble
after the Installation of the ACS and the NICMOS Cooling System, ed. S.
Arribas, A. Koekemoer, \& B. Whitmore (Baltimore: Space Telescope Science Institute), 373\\
McArthur, B. E., et al. 2001, ApJ, 560, 907\\
McLaughlin, D. B. 1945, PASP, 57, 69\\
McLaughlin, D. B. 1940, ApJ, 91, 369\\
McLaughlin, D. B. 1939, Popular Astronomy, 47, 410\\
McLaughlin, D. B. 1937, {\it Publ. Obs. Univ. Mich.}, 6, No. 12, 107\\
Morales-Rueda, L., Still, M. D., Roche, P., Wood, J. H., \& Lockley, J. J. 2002,
MNRAS, 329, 597\\
Nelan, E., et al. 2011, ``Fine Guidance Sensor Instrument Handbook'', Version
19.0, (Balitmore: STScI)\\
Nikitin, S. A., Vshivkov, V. A., \& Snytnikov, V. N. 2000, Astronomy Letters, 26, 362\\
O'Brien, T. J., \& Bode, M. F. 2008, in ``Classical Novae'', ed. M. F.
Bode and A. Evans (Cambridge University Press: New York), p285\\ 
Payne-Gaposchkin, C. 1957, {\it The Galactic Novae} (Dover, New York)\\
Pek\"{o}n, Y., \& Balman, \c{S}. 2008, MNRAS, 388, 921\\
Perrine, C. D. 1902, ApJ, 14, 249\\
Perryman, M. A. C., et al. 1997, A\&A, 323, 49\\
Rieke, G. H., Lebofsky, M. J. 1985, ApJ, 288, 618\\
Roeser, S., Demleitner, M., \& Schilback, E. 2010, AJ, 139, 2440\\
Seaquist, E. R., Bode, M. F., Frail, D. A., Roberts, J. A., Evans, A., \&
Albinson, J. S. 1989, ApJ, 344, 805\\
Shara, M. M., Zurek, D., De Marco, O., Mizusawa, T., Williams, R., \& Livio, M.
2012, ApJ, 143, 143\\
Sherrington, M. R., \& Jameson, R. F., 1983, MNRAS, 205, 265\\
Soderblom, D. R., Nelan, E., Benedict, G. G., McArthur, B., Ramirez, I.,
Spiesman, W., \& Jones, B. F. 2005, AJ, 129, 1616\\
Spencer Jones, H. 1931, Cape Ann., 10, 9\\
Standish, E. M. Jr. 1990, A\&A, 233, 252\\
Starrfield, S., Iliadis, C., Hix, W. R., Timmes, F. X., \& Sparks, W. M. 2009,
ApJ, 692, 1532\\
Strope, R. J., Schaefer, B. E., \& Henden, A. A. 2010, ApJ, 140, 34\\
Townsley, D. M., \& Bildsten, L. 2004, ApJ, 600, 390\\
van den Bos, W. H., \& Finsen, W. S. 1931, MNRAS, 92, 19\\
van den Bergh, \& Pritchet, C. J. 1986, PASP, 98, 110\\
van Leeuwen, F. 2007, in ``Hipparcos, the new reduction of the raw data
(Dordrecht: Springer)\\
Verbunt, F. 1987, A\&A Supp., 71, 339\\
Wade, R. A., Harlow, J. J. B., \& Ciardullo, R. 2000, PASP, 112, 614\\
Warner, B. 2008, in ``Classical Novae'', ed. M. F.
Bode and A. Evans (Cambridge University Press: New York), p16\\
Weaver, H. 1974, in Contopolous G., ed., Highlights of Astronomy, Vol. 3.,
(Reidel: Dordrecht), p. 509\\
Williams, R. E., \& Gallagher, J. S. 1979, ApJ, 228, 482\\
Wu, C. -C., Holm, A. V., Panek, R. J., Raymond, J. C., Harmann, L. W., \&
Swank, J. H. 1989, ApJ, 339, 443\\
Wyse, A. B. 1940, Pub. Lick. Obs., 14, 93\\
Yamashita, Y., Nariai, K., \& Normoto, Y. 1978,  ``An Atlas of
Representative Stellar Spectra'', (Halsted Press: New York)\\
\end{flushleft}

\begin{sidewaystable}
\footnotesize
\begin{tabular}{cccccccccc}
\multicolumn{10}{c}{Table 1. Astrometric Reference Stars\footnote{\scriptsize
The $\alpha_{\rm 2000}$ and $\delta_{\rm 2000}$ are GSC2 coordinates and have the following
epochs: V603 Aql = 1990.63, DQ Her = 1991.68, GK Per = 1989.76, and RR Pic =
1995.07.}}\\
\hline
\hline
ID&$\alpha_{\rm 2000}$ & $\delta_{\rm 2000}$&$\mu_{\alpha}$ (mas yr$^{\rm -1}$)&$\mu_{\delta}$ (mas yr$^{\rm -1}$)&Sp. Ty.& V & $(B -V)$&A$_{\rm V}$& $\pi$ (mas)\footnote{\scriptsize As noted in the text, the error bars on the reference 
star parallaxes listed here result from astrometric solutions to the various
FGS data sets, and are not independent measurements. For 
distant reference stars, the final errors on the 
parallaxes reported by the astrometric solution are more heavily weighted
by those of the input spectroscopic parallaxes. For nearer stars, the error 
in the parallaxes are more heavily weighted by the positional uncertainty of 
the FGS measurements. An example of the latter is V603 Aql R02. The input 
spectroscopic parallax for this star was 6.71 $\pm$ 1.4 mas. The error
bar on the spectroscopic parallax for V603 Aql R02 is larger than the
typical positional precision possible with the FGS. Thus, the final error bar 
on the parallax from the astrometric solution for this nearby star is 
dominated by the precision of the FGS measurements. In contrast, for a 
distant reference star such as V603 Aql R07, the error on the input 
spectroscopic parallax was smaller than that of the intrinsic measurement 
error of the FGS. Thus, for this star, GaussFit assigns a higher weight to the input spectroscopic parallax.
} \\
\hline
V603 Aql R01&18:49:19.30&+00:34:26.30&0.40$\pm$0.244&$-$21.81$\pm$0.17&G2V &13.36&1.05& 1.39& 3.28$\pm$0.13\\
V603 Aql R02&18:49:01.90&+00:33:46.40&$-$9.48$\pm$0.39&$-$30.17$\pm$0.23&K5V &13.62&1.16& 0.29&6.76$\pm$0.25\\
V603 Aql R03&18:48:48.20&+00:34:59.80&3.13$\pm$0.19&$-$4.59$\pm$0.20&F0V &15.78&1.45& 3.16&0.90$\pm$0.05\\
V603 Aql R04&18:48:46.30&+00:35:50.60&4.95$\pm$0.15&$-$7.97$\pm$0.16&K2III &14.17&2.31& 3.56& 0.92$\pm$0.05\\
V603 Aql R05&18:48:41.00&+00:38:49.50&6.45$\pm$0.32&$-$7.45$\pm$0.25&F0V & 9.88&0.39& 0.29&3.56$\pm$0.15\\
V603 Aql R06&18:48:49.50&+00:38:26.80&25.18$\pm$0.43&$-$26.84$\pm$0.43& K0V &12.43&0.82& 0.00&4.79$\pm$0.24\\
V603 Aql R07&18:48:55.80&+00:36:28.80&$-$2.31$\pm$0.22&$-$4.60$\pm$0.21&K5III&15.38&2.26& 2.32&0.23$\pm$0.01\\
V603 Aql R08&18:48:53.00&+00:36:18.20&2.79$\pm$0.18&$-$0.29$\pm$0.20&K0V &14.98&1.02& 0.91& 2.48$\pm$0.167\\
V603 Aql R09&18:49:00.79&+00:36:37.60&$-$3.59$\pm$0.33&$-$0.64$\pm$0.27&B0V&15.00&1.70& 6.20& 0.27$\pm$0.01\\
DQ Her R01\footnote{\scriptsize The spectroscopic parallax for this object is $\pi$ = 3.02 mas, but
the astrometric solution suggests that it is further away: $\pi$ = 1.85 mas. It is quite
likely that this is an unresolved binary star with both components having similar spectral types.}  &18:07:38.42&+45:47:35.40&2.50$\pm$0.15&$-$3.37$\pm$0.17&F2V&10.80&0.52&0.10&1.82$\pm$0.15\\
DQ Her R02  &18:07:33.18&+45:47:30.60&$-$5.32$\pm$0.33&7.68$\pm$0.38&G1V   &13.05&0.66&0.05&1.87$\pm$0.20 \\
DQ Her R03  &18:07:26.70&+45:47:57.50&$-$3.24$\pm$0.37&$-$3.54$\pm$0.32&K3III &11.57&0.66&0.14&0.63$\pm$0.07 \\
DQ Her R04  &18:07:23.90&+45:49:47.60&$-$13.87$\pm$0.34&$-$13.26$\pm$0.36&G2V   &14.23&0.66&0.10& 1.20$\pm$0.11 \\
DQ Her R05  &18:07:29.97&+45:49:49.70&3.34$\pm$0.33&$-$8.13$\pm$0.38&G0V   &15.12&0.63&0.11 &0.70$\pm$0.06 \\
DQ Her R06  &18:07:20.44&+45:51:14.80&$-$12.64$\pm$0.35&$-$25.45$\pm$0.41&K4V   &14.30 &1.10&0.10&3.78$\pm$0.27 \\
DQ Her R07  &18:07:17.75&+45:52:53.70&$-$0.79$\pm$0.31&6.58$\pm$0.35&G2V   &14.65&0.63&0.13&0.93$\pm$0.14 \\
DQ Her R08  &18:07:25.45&+45:54:06.50&$-$4.18$\pm$0.32&$-$16.21$\pm$0.35&G2V   &14.35&0.67&0.13&1.28$\pm$0.14 \\
DQ Her R09  &18:07:13.60&+45:55:26.60&$-$10.00$\pm$0.30&$-$4.91$\pm$0.34&F8V   &12.71&0.58&0.09&1.59$\pm$0.18 \\
\hline
\end{tabular}
\end{sidewaystable}
\clearpage
\begin{sidewaystable}
\footnotesize
\begin{tabular}{cccccccccc}
\multicolumn{10}{c}{Table 1. Astrometric Reference Stars (cont.)}\\
\hline
\hline
ID&$\alpha_{\rm 2000}$& $\delta_{\rm 2000}$&$\mu_{\alpha}$ (mas yr$^{\rm -1}$)&$\mu_{\delta}$ (mas yr$^{\rm -1}$)&Sp. Ty.& V & $(B -V)$&A$_{\rm V}$& $\pi$ (mas)\\
\hline
GK Per R01  &03:31:14.18&+43:54:24.60&54.17$\pm$0.35&$-$4.36$\pm$0.36&M0V&15.29&1.38&0.0&3.53$\pm$0.19\\
GK Per R02  &03:31:24.08&+43:54:43.40&1.78$\pm$0.34&$-$1.39$\pm$0.31&F5V&14.49&0.69&0.85&0.87$\pm$0.06\\
GK Per R04  &03:31:22.38&+43:55:26.40&$-$1.84$\pm$0.30&$-$24.83$\pm$0.24&G6V&14.84&1.04&1.32&2.07$\pm$0.12\\
GK Per R07  &03:31:32.48&+43:55:52.10&10.45$\pm$0.16&$-$15.32$\pm$0.16&F7V&13.18&0.75&1.08&2.12$\pm$0.10\\
GK Per R08  &03:31:03.99&+43:53:19.70&$-$1.86$\pm$0.25&$-$13.14$\pm$0.24&K1IV&15.78&1.21&0.92&1.87$\pm$0.11\\
RR Pic R01 &06:36:27.60&$-$62:39:04.80&$-$16.34$\pm$0.48&13.77$\pm$0.44&G1V&12.27&0.60&0.00&2.66$\pm$0.22\\
RR Pic R03 &06:36:07.26&$-$62:39:34.80&$-$8.08$\pm$0.77&3.24$\pm$0.77&G1V&14.71&0.76&0.18&0.89$\pm$0.07\\
RR Pic R04 &06:35:51.31&$-$62:37:46.20&0.29$\pm$0.36&11.36$\pm$0.38&F5V&13.98&0.63&0.29&0.82$\pm$0.07\\
RR Pic R05 &06:35:28.84&$-$62:38:34.30&$-$2.46$\pm$0.39&15.64$\pm$0.31&G2V&15.13&0.88&0.47&0.95$\pm$0.07\\
RR Pic R06 &06:35:40.34&$-$62:38:41.40&1.87$\pm$0.51&0.73$\pm$0.39&G5V&15.11&0.86&0.22&0.96$\pm$0.09\\
RR Pic R07 &06:35:10.83&$-$62:37:49.10&2.33$\pm$0.76&15.45$\pm$0.62&G5V&15.22&0.80&0.20&1.00$\pm$0.09\\
RR Pic R08 &06:34:57.10&$-$62:37:11.70&6.16$\pm$0.33&12.38$\pm$0.30&K0IV&13.93&1.17&0.81&0.84$\pm$0.06\\
RR Pic R09 &06:34:50.11&$-$62:37:40.40&3.12$\pm$0.41&6.61$\pm$0.36&F6V&15.05&0.61&0.22&0.55$\pm$0.05\\
RR Pic R10 &06:35:01.15&$-$62:38:14.50&1.89$\pm$0.76&16.33$\pm$0.64&K1.5V&15.28&0.96&0.10&1.82$\pm$0.19\\
RR Pic R11 &06:35:07.93&$-$62:39:53.20&$-$0.20$\pm$0.40&21.07$\pm$0.38&G9IV&13.03&0.88&0.26&1.01$\pm$0.08\\
RR Pic R12 &06:36:25.98&$-$62:38:12.50&$-$3.21$\pm$0.27&13.92$\pm$0.21&G5III&10.41&0.85&0.17&2.60$\pm$0.14\\
\hline
\end{tabular}
\end{sidewaystable}
\clearpage

\begin{deluxetable}{ccccccc}
\renewcommand\thetable{2}
\tablecolumns{7}
\tablewidth{0pt}
\centering
\tablecaption{Astrometric Properties of Program Classical Novae\tablenotemark{c}}
\tablehead{\colhead{Nova}&\colhead{$\alpha_{\rm 2000}$} & \colhead{$\delta_{\rm 2000}$}& \colhead{$\mu_{\alpha}$} & 
\colhead{$\mu_{\delta}$} & \colhead{Parallax} & \colhead{LKH}\\
\colhead{}&\colhead{}&\colhead{}&\colhead{(mas yr$^{\rm -1}$)}&\colhead{(mas yr$^{\rm -1}$)}&\colhead{(mas)}& \colhead{(mag)}}
\startdata
V603 Aql & 18:48:54.64 & $+$00:35:02.9 & 11.916 $\pm$ 0.142& $-$10.240 $\pm$ 0.131 &
4.011 $\pm$ 0.137 & $-$0.01 \\
DQ Her   & 18:07:30.26 & $+$45:51:32.1 & $-$2.125 $\pm$ 0.210&13.301 $\pm$ 0.244 & 2.594 $\pm$ 0.207 & $-$0.05 \\
GK Per   & 03:31:12.01 & $+$43:54:15.4 & $-$6.015 $\pm$ 0.197& $-$22.767 $\pm$ 0.208&
2.097 $\pm$ 0.116 & $-$0.08 \\
RR Pic   & 06:35:36.07 & $-$62:38:24.3 & 5.204 $\pm$ 0.222 & 0.245 $\pm$ 0.281 & 1.920 $\pm$ 0.182 & $-$0.07 \\
\enddata
\tablenotetext{c}{See footnote $a$ for Table 1.}
\end{deluxetable}
\clearpage

\begin{deluxetable}{ccc}
\renewcommand\thetable{3}
\tablecolumns{3}
\tablewidth{0pt}
\centering
\tablecaption{Astrometric Distances versus Nebular Expansion Parallax Distances\tablenotemark{d}}
\tablehead{
\colhead{Nova} &\colhead{Astrometric Distance} & \colhead{Nebular Distance}} 
\startdata
V603 Aql & 249$^{\rm +9}_{\rm -8}$ & 328$^{\rm +60}_{\rm -29}$\\
DQ Her & 386$^{\rm +33}_{\rm -29}$ & 545$^{\rm +81}_{\rm -70}$\\
GK Per & 477$^{\rm +28}_{\rm -25}$ & 460$^{\rm +69}_{\rm -59}$\\
RR Pic & 521$^{\rm +54}_{\rm -45}$ & 580$^{\rm +89}_{\rm -73}$\\
\enddata
\tablenotetext{d}{From Downes \& Duerbeck (2000).}
\end{deluxetable}
\clearpage

\begin{deluxetable}{ccccccc}
\renewcommand\thetable{4}
\tablecolumns{7}
\tablewidth{0pt}
\centering
\tablecaption{Outburst Details for Program Classical Novae}
\tablehead{
\colhead{Nova} &\colhead{Year of Maximum} & \colhead{$t_{\rm 2}$} & 
\colhead{$t_{\rm 3}$} & \colhead{V$_{\rm Max}$} & \colhead{$v_{\rm principal}$} &
\colhead{M$_{\rm V_{\rm max}}$}\tablenotemark{e}}

\startdata
V603 Aql &1918.764 &  4.5 $\pm$ 0.5 &  9.3 $\pm$ 2.3 & $-$1.4 $\pm$ 0.3: & 1600 & 
$-$8.60$^{\rm -0.08}_{\rm +0.07}$\\
DQ Her   &1934.978 & 70.0 $\pm$ 5.0 & 95.6 $\pm$ 3.0 &  1.3 $\pm$ 0.2 &  315  &
$-$6.94$^{\rm -0.18}_{\rm +0.17}$\\
GK Per   &1901.148 &  6.3 $\pm$ 0.6 & 13.0 $\pm$ 0.0 &  0.0 $\pm$ 0.1 & 1200  &
$-$9.35$^{\rm -0.13}_{\rm +0.11}$\\
RR Pic   &1925.436 & 78.3 $\pm$ 4.7 &136.0 $\pm$ 13.2&  1.1 $\pm$ 0.1 &  310  &
$-$7.61$^{\rm -0.22}_{\rm +0.19}$\\
\enddata
\tablenotetext{e}{The errors on these absolute magnitudes represent only the 
uncertainty due to the parallax, and do not include the uncertainty in V$_{\rm Max}$.}
\end{deluxetable}
\clearpage

\begin{figure}
\epsscale{1.00}
\plotone{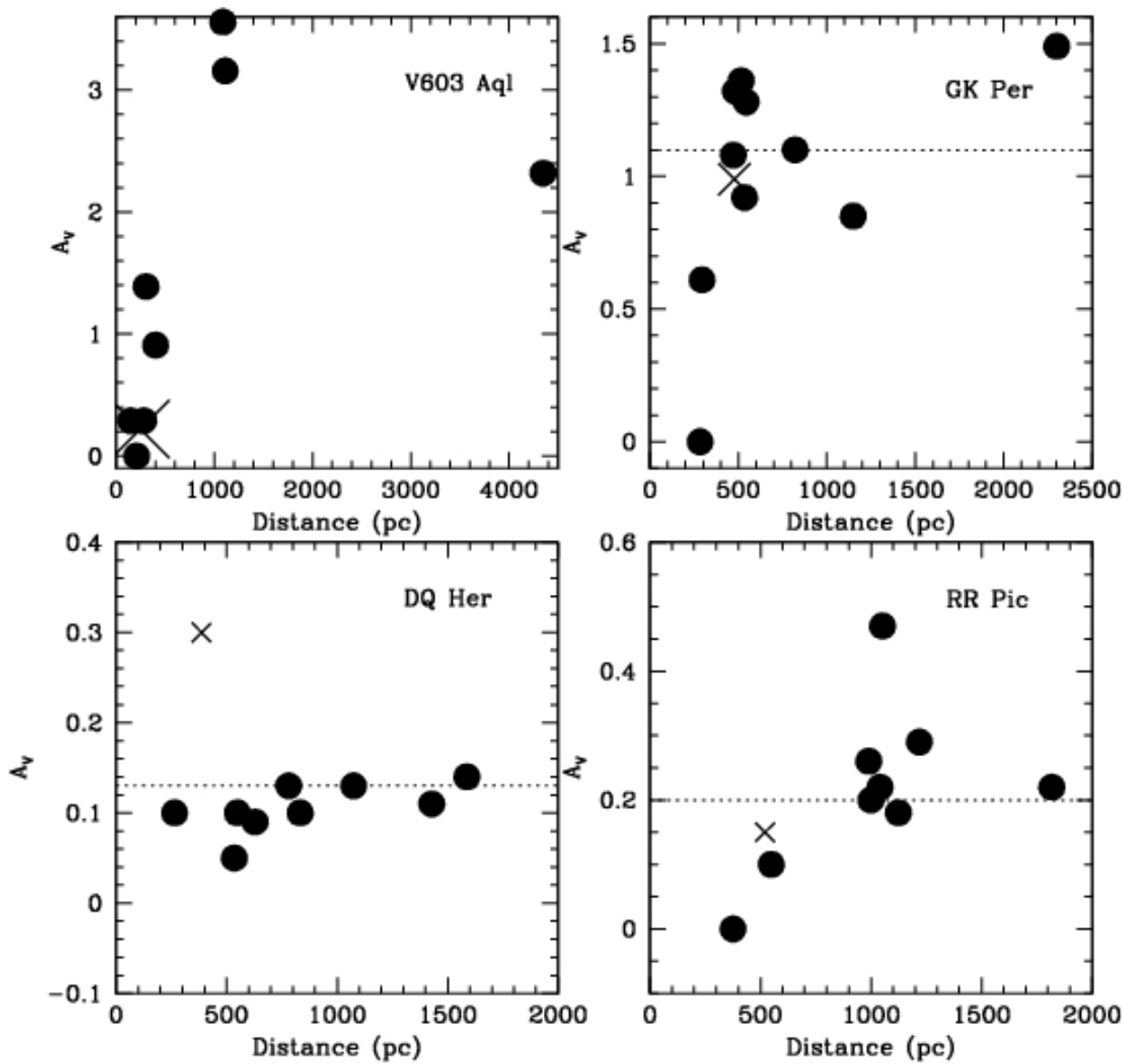}
\caption{The derived extinction of the astrometric reference stars in the four CNe
fields versus their distances derived from their spectroscopic parallaxes. The 
positions of the classical novae are indicated with an ``$\times$''. The mean 
(IRSA) line-of-sight extinction (averaged over a 2$^{\circ}$ field-of-view) of 
each field is represented by a horizontal dotted line 
(except for V603 Aql, where the line-of-sight extinction at its low
galactic latitude is A$_{\rm V}$ $\geq$ 15 mag). For GK Per we have plotted
our results for all ten of the reference stars, not just the five used in the
astrometric solution.}

\label{figure1}
\end{figure}

\begin{figure}
\epsscale{1.00}
\plotone{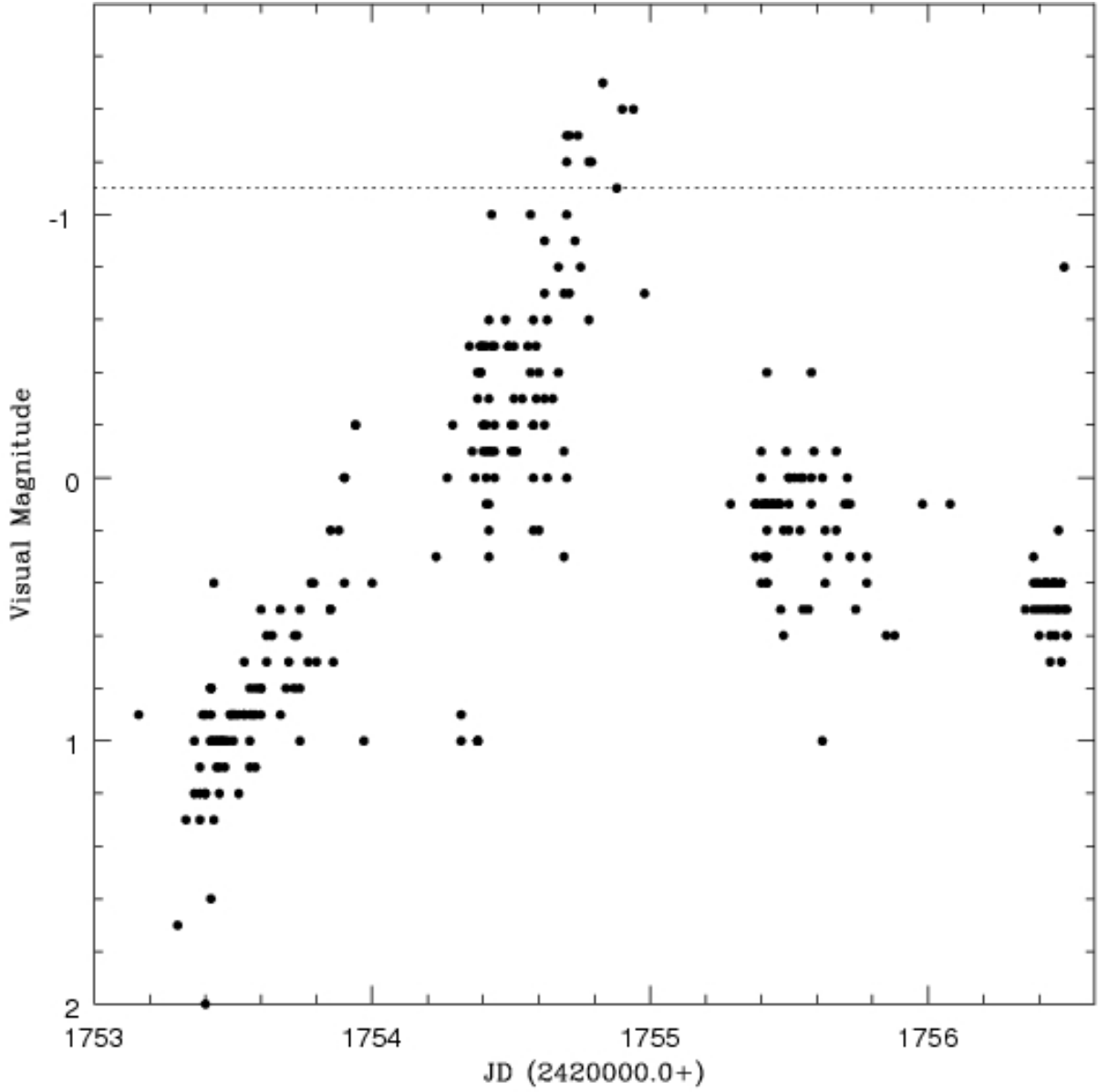}
\caption{The light curve of V603 Aql near visual maximum
(Campbell 1919). The dashed line at m$_{\rm V}$ = $-$1.1 demarcates the commonly 
quoted value for its visual maximum magnitude.}
\label{figure2}
\end{figure}

\begin{figure}
\epsscale{1.00}
\plotone{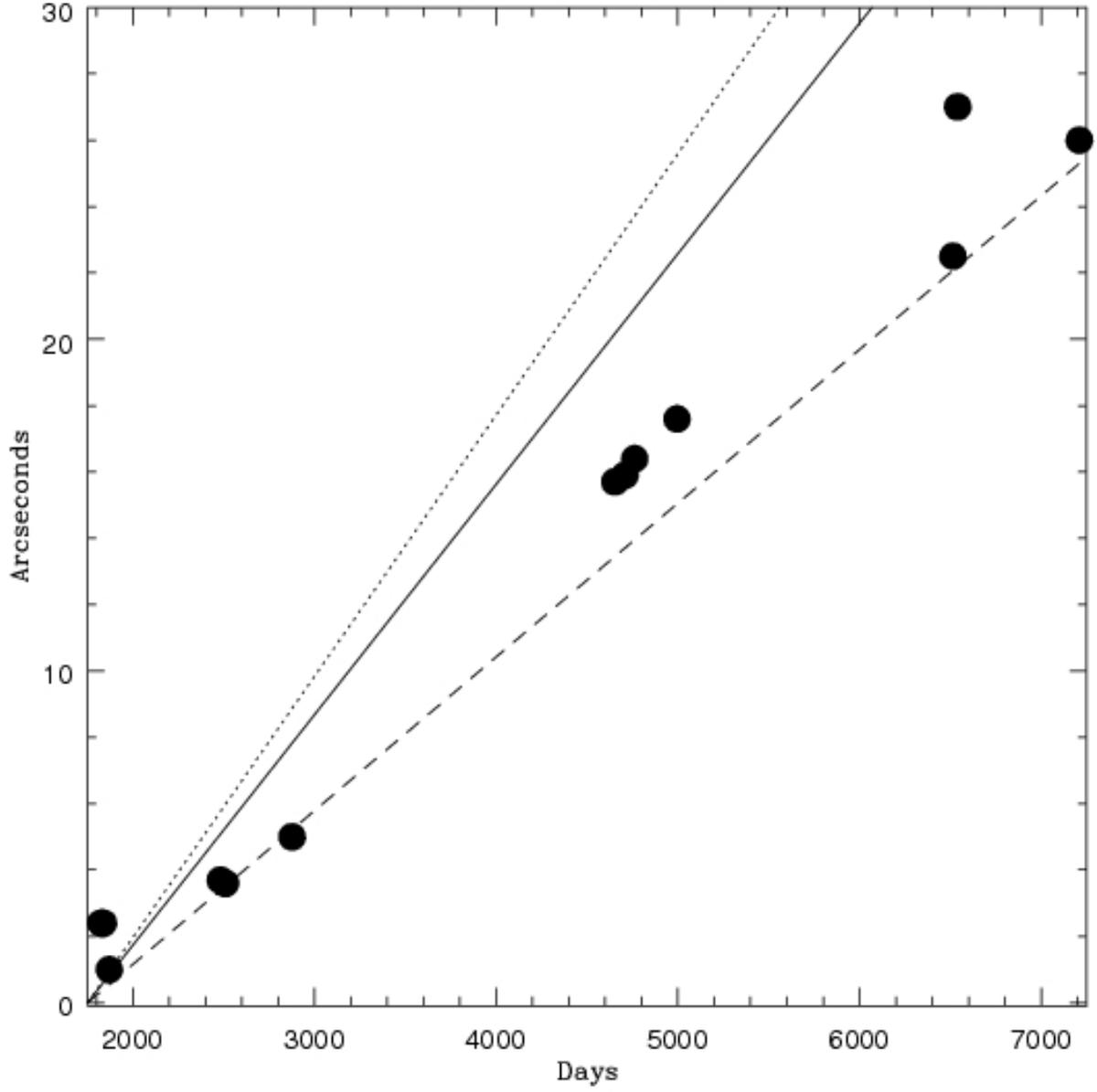}
\caption{The angular expansion rate of the nebular shell of V603 Aql, data (solid
circles) from
Wyse (1940). The dashed line is the expansion rate if the ejecta velocity
was 1,000 km s$^{\rm -1}$, the solid line is for 1,500 km s$^{\rm -1}$, and
the dotted line is 1,700 km s$^{\rm -1}$.}
\label{figure3}
\end{figure}

\begin{figure}
\epsscale{1.00}
\plotone{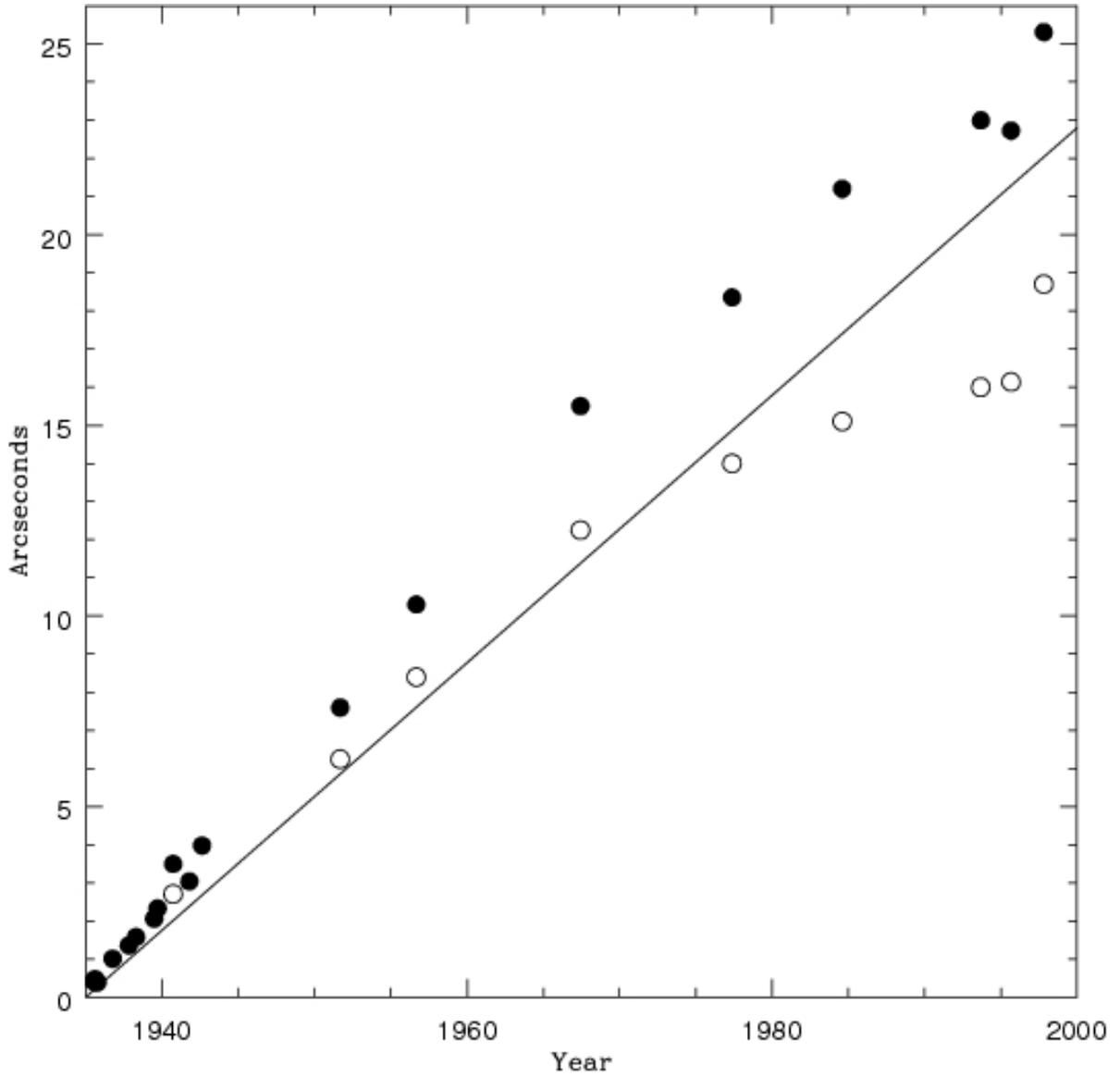}
\caption{The angular expansion rate of the nebular shell of DQ Her. Solid
circles are the measurements of the diameter of major axis, while open circles 
indicate the diameter of the minor axis. The solid line is the projected angular 
size of a shell that was ejected with v$_{\rm exp}$ = 320 km s$^{\rm -1}$ at the 
time of outburst.}
\label{figure4}
\end{figure}

\begin{figure}
\epsscale{1.00}
\plotone{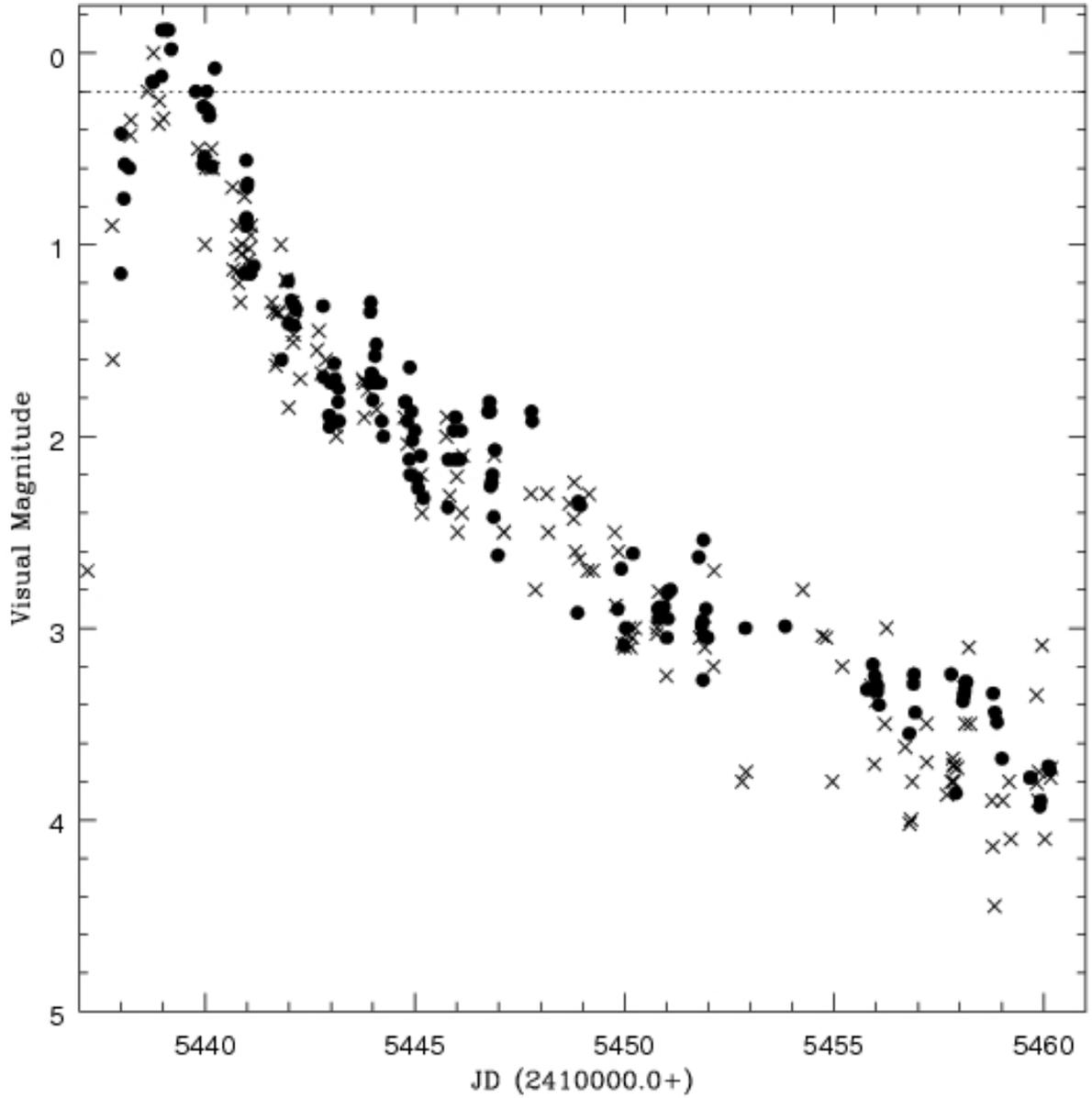}
\caption{The light curve of GK Per near visual maximum using data from
Campbell (1903). The solid circles are measurements where we have used
modern $V$ magnitudes for the comparison stars to recalibrate the early portion
of the light curve of GK Per. The crosses are 
for data taken directly from Campbell. The dotted line at 
m$_{\rm V}$ = 0.2 is the commonly quoted value for its visual maximum.}
\label{figure5}
\end{figure}

\renewcommand{\thefigure}{6}
\begin{figure}
\epsscale{1.00}
\plotone{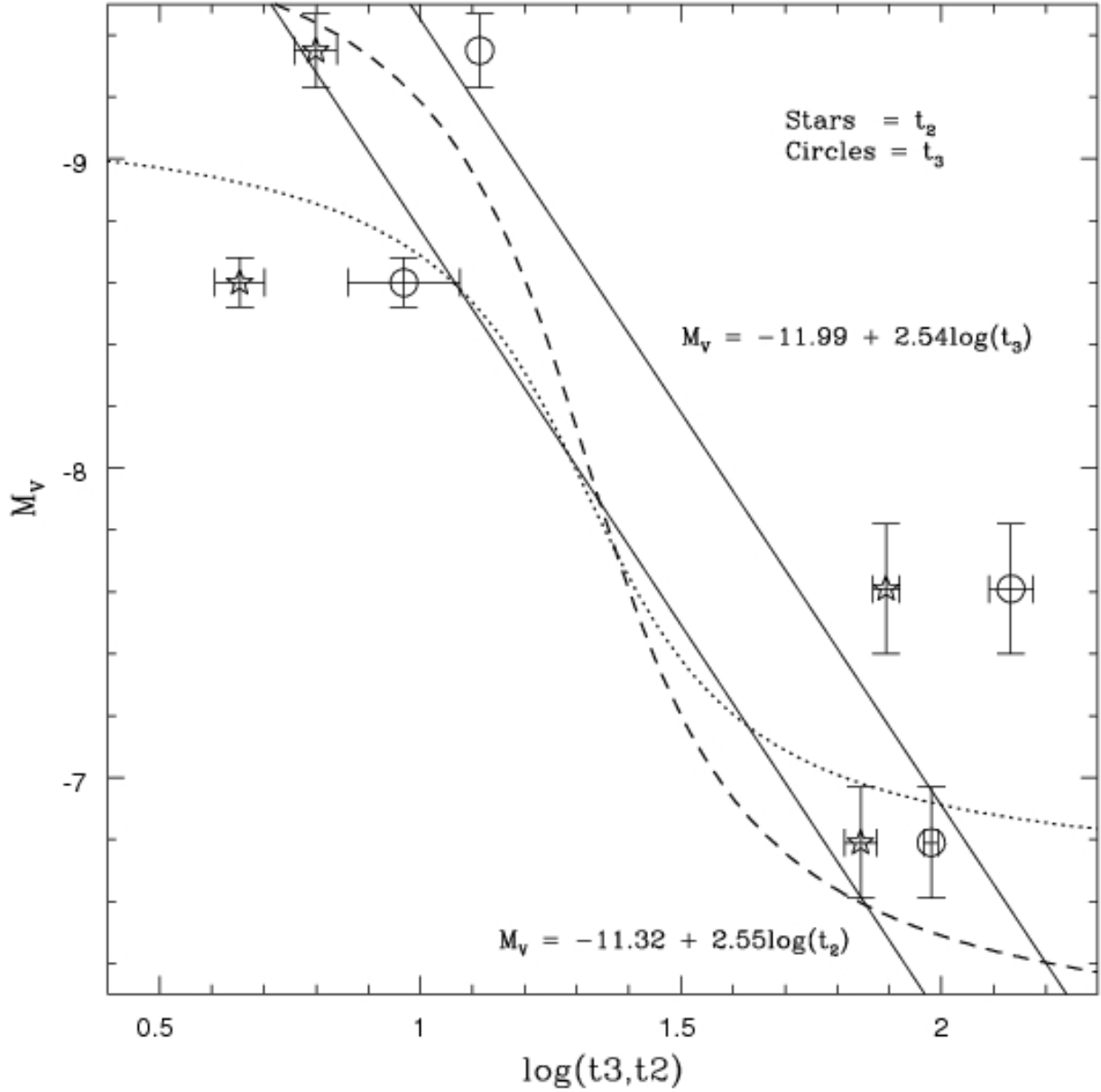}
\caption{The Maximum Magnitude-Rate of Decline plot for the program
novae. The linear relationships for $t_{\rm 2}$ and $t_{\rm 3}$
from Downes \& Duerbeck (2000) are plotted as solid lines and labeled. The dotted 
line is the Della Valle \& Livio (1995) arctangent relationship for $t_{\rm 2}$, 
and the dashed line is this law as updated by Downes \& Duerbeck. The error
bars on the absolute magnitudes are those due to the error in the parallax,
and do not include the uncertainty in the peak visual magnitudes of the CNe.}
\label{figure6}
\end{figure}

\end{document}